# P-HGRMS: A Parallel Hypergraph Based Root Mean Square Algorithm for Image Denoising

| Tejaswi Agarwal | Saurabh Jha | B. Rajesh Kanna |
|---|---|---|
| Undergraduate Student | Undergraduate Student | Advisor |
| VIT University, Chennai, India | VIT University, Chennai, India | VIT University, Chennai, India |
| tejaswi.agarwal2010@vit.ac.in | saurabh.jha2010@vit.ac.in | rajeshkanna.b@vit.ac.in |

**ABSTRACT**

This paper presents a parallel Salt and Pepper (SP) noise removal algorithm in a grey level digital image based on the Hypergraph Based Root Mean Square (HGRMS) approach. HGRMS is generic algorithm for identifying noisy pixels in any digital image using a two level hierarchical serial approach. However, for SP noise removal, we reduce this algorithm to a parallel model by introducing a cardinality matrix and an iteration factor, *k*, which helps us reduce the dependencies in the existing approach. We also observe that the performance of the serial implementation is better on smaller images, but once the threshold is achieved in terms of image resolution, its computational complexity increases drastically. We test P-HGRMS using standard images from the Berkeley Segmentation dataset on NVIDIAs Compute Unified Device Architecture (CUDA) for noise identification and attenuation. We also compare the noise removal efficiency of the proposed algorithm using Peak Signal to Noise Ratio (PSNR) to the existing approach. P-HGRMS maintains the noise removal efficiency and outperforms its sequential counterpart by 6 to 18 times (6x - 18x) in computational efficiency.

**Categories and Subject Descriptors**: I.3 [**Computer Graphics**]: Hardware Architecture –*Graphics Processor, Parallel Processing*

## 1. INTRODUCTION

To overcome limitations in the existing spatial image filtering and noise removal algorithms, Kanna [1] proposed the HGRMS technique which uses Hypergraph based Root Mean Square approach for Salt and Pepper (SP) noise removal. HGRMS is a generic approach, which models the input image using Image Neighborhood hypergraph (INHG) parameters alpha, beta (α, β) to identify the noisy pixels. The identification of noisy hyperedge (NH) involves two sequential operations such that the NH should satisfy the property of isolated hyperedge and its cardinality equal to one. This causes severe performance penalties in computation.

## 2. PROPOSED ALGORITHM

Our noise reduction algorithm works on the following criterion: binary classification of hyperedges of an image (H0 noisy hyperedge and H1 no noisy hyperedge) and filtering the noisy parts. To modify the existing algorithm we propose the use of a cardinality matrix, and updating the image to remove the noise based on the values in the cardinality matrix. The cardinality matrix is generated based on the alpha, beta (α, β) neighborhood of the pixels in an image. The cardinality matrix serves as a platform to exploit parallelism using CUDA threads to update the original image with the Root Mean Square value of its corresponding beta neighborhood pixels. We preserve the image adaptive nature of the existing algorithm by introducing a new parameter, known as iteration factor, *k*. In our approach, to parallelize the algorithm, we iterate over the modified pixel values until there is no alpha-beta noise remaining in the image based on the iteration factor.

Algorithm 1 illustrated below is used to form the cardinality matrix of the image I. The cardinality matrix is formed by finding the cardinality of each pixel, i.e. counting the number of closed neighborhood alpha-beta pixels with respect to that pixel in I. The cardinality number of each pixel is saved in another matrix, which we call the cardinality matrix, C.

Each CUDA thread computes the cardinality value of a pixel in image in parallel with other CUDA threads depending on its threadIdx and blockIdx value. Thus Cardinality matrix, C is generated in parallel.

---
**Algorithm 1:** findIsolatedPixels to generate cardinality matrix

**findIsolatedPixels**(*d_image, *d_card, col, row)
1. **do in parallel** on each thread
2.     c := blockDim.x*blockIdx.x+threadIdx.x
3.     r := blockDim.y*blockIdx.y+threadIdx.y
4.     **if** r<row and c<col
5.         **for** i:=r-beta to r+beta+1
6.             **for** j:=c-beta to c+beta+1
7.     **if** i<row and i>=0 and j<col and j>=0
8.         **if** d_image[i*col+j] < d_image[r*col+c]+alpha and d_image[i*col+j]>d_image[r*col+c]-alpha
9.             d_card[i*col+j]:= d_card + 1
10. **end do**
11. cudaThreadSynchronize()

---
**Algorithm 2:** Remove noise for image enhancement

**removeNoise**(*d_image,*d_tempImage, *d_card, col, row)
1. j=blockDim.x*blockIdx.x+threadIdx.x;
2. i=blockDim.y*blockIdx.y+threadIdx.y;
3. **if** i<row and j<col
4.     sum:=0
5.     pix_count:=0
6.     flag:=0
7.     **if** d_card[i*col+j]<=2
8.         **for** i1:=i-beta to i1<i+beta+1
9.             **for** j1:=j-beta to j1<j+beta+1
10.                 pix_count = pix_count + 1
11.             **if** i1<row and i1>=0 and j1<col and j1>=0

```
12.  if not (d_image[i1*col+j1] < d_image[i*col+j]+alpha+1 and
              d_image[i1*col+j1]>d_image[i*col+j]-alpha
             )
13.     sum = sum + d_image[i1*col+j1] * d_image[i1*col+j1]
14.                        flag = flag + 1
15.        if flag>pix_count-3
16.               d_image'[i*col+j]=sum/flag
17. cudaThreadSynchronize()
```

In algorithm 2, for each cell in C, if cardinality is found to be less than the threshold value, 3, we find the open neighborhood alpha-beta of the pixel in I, and count the number of pixels which exceed this value. If the count exceeds threshold value, we replace identified noisy pixel by Root Mean Square value of open neighborhood of this pixel in a temporary image matrix, I'. The temporary matrix enables us to remove data race conditions. Once all the CUDA threads are synchronized, we copy the temporary image matrix to original image matrix. We repeat the above steps until no further noise remains with respect to alpha and beta. We iteratively call algorithm 1 and algorithm 2 one after another until no further noise remains with respect to alpha and beta.

The iteration factor, $k$, helps us to control the number of iterations. The number of times algorithm 1 and algorithm 2 are called in succession depends on this iteration factor, k. This factor is necessary to preserve the image adaptive nature of HGRMS algorithm. The value of k is dependent on the INHG parameter, $\beta$. For our computation, we use $k=5$.

## 3. EXPERIMENTAL RESULTS

The existing algorithm HGRMS was implemented and tested on a system with an Intel Core i5 Processor with 4 GB of RAM. For the GPU implementation we used and Amazon AWS cluster, having 2 X NVIDIA Tesla M2050 GPUs having 1690 GB of instance storage and 22 GiB of memory. The images are acquired from the Berkeley Segmentation dataset [2].

Table 1: Execution time in milliseconds to various levels of noise

| Noise | 1.png (Time in ms) | | 2.png (Time in ms) | | 3.png (Time in ms) | | 4.png (Time in ms) | | 5.png (Time in ms) | |
|---|---|---|---|---|---|---|---|---|---|---|
|  | CPU | GPU | CPU | GPU | CPU | GPU | CPU | GPU | CPU | GPU |
| 5% | 14 | 2.23 | 14 | 2.21 | 17 | 1.91 | 14 | 1.81 | 14 | 1.92 |
| 10% | 15 | 2.03 | 14 | 1.86 | 14 | 1.95 | 16 | 1.91 | 15 | 1.94 |
| 15% | 14 | 2.1 | 16 | 1.90 | 15 | 2.01 | 15 | 1.95 | 16 | 2.00 |
| 20% | 16 | 2.12 | 17 | 1.99 | 16 | 2.04 | 16 | 2.00 | 16 | 2.01 |

Table 1 shown above gives us the run time of both algorithms with image noise levels ranging from 5 % to 20 %. It is evident from the results that a change in noise level does not significantly impact the performance of the proposed algorithm. This was expected as our algorithm is dependent on the size of the image, irrespective of any other factor.

Table 2: PSNR values of HGRMS and PHGRMS with different levels of noise.

| Noise / Image | 1.png | | 2.png | | 3.png | | 4.png | | 5.png | |
|---|---|---|---|---|---|---|---|---|---|---|
|  | PHGRMS | HGRMS | PHGRMS | HGRMS | PHGRMS | HGRMS | PHGRMS | HGRMS | PHGRMS | HGRMS |
| 5% | 26.94 | 26.97 | 29.12 | 29.15 | 31.70 | 31.69 | 29.76 | 29.91 | 28.97 | 28.96 |
| 10% | 24.60 | 24.55 | 30.78 | 30.76 | 27.96 | 27.91 | 26.82 | 26.83 | 26.11 | 26.06 |
| 15% | 22.17 | 22.11 | 25.89 | 25.86 | 24.30 | 24.31 | 23.64 | 23.63 | 23.46 | 23.39 |
| 20% | 19.67 | 19.64 | 22.45 | 22.40 | 21.19 | 21.18 | 20.66 | 20.62 | 20.76 | 20.74 |

Restoration performance is quantitatively measured by the peak signal-to-noise ratio (PSNR). Table 2 shown above compares PSNR values of the existing HGRMS algorithm and the proposed P-HGRMS algorithm.

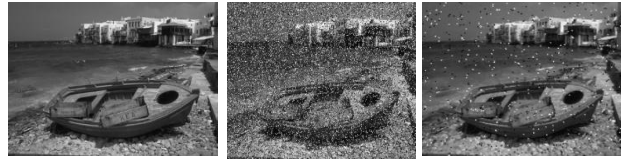

**Figure 1**: (a) Original Image (b) 20% noise (c) Restored image

Figure 1 shows one sample (5.png) of the total five images used for our analysis. The restored image (c) is formed using our proposed PHGRMS algorithm.

As noted below, even with a small image size of 128 X 128 we observe a speed up of 6x. The parallel CUDA implementation is 18x faster with an image resolution of 2048x2048.

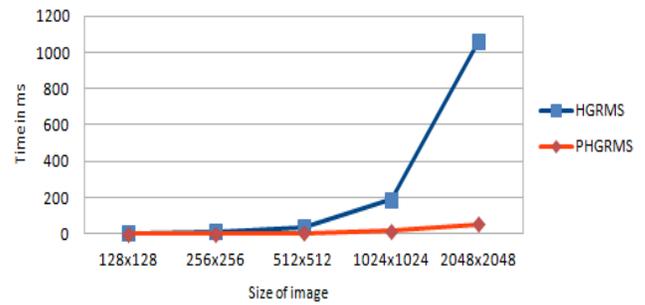

**Figure 2**: Speedup of P-HGRMS compared to HGRMS

## 4. CONCLUSION:

In this paper, we proposed a two phase parallel algorithm for noise filtering in images with preserving the image details. Our results show that the proposed algorithm outperforms its existing Hypergraph based root mean square approach. It is worth mentioning that when images are contaminated with high SP noise, the proposed scheme is able to clean images quiet efficiently without loss of image details. The proposed algorithm in this work is highly performance efficient as compared to its serial implementation. Even at higher noise ranges of 20 % the algorithm performed as efficiently as its serial counterpart.

This algorithm represents a very fast and promising approach in denoising salt and pepper noise in digital images. Future work includes reducing processing time further by testing with the MPI implementation and superior hardware capabilities. We also plan to devise a mechanism where the algorithm would itself decide the iteration factor, $k$, based on the input image, rather than a pre-defined iteration factor.